# Low-Complexity Acoustic Scene Classification Using Parallel Attention-Convolution Network

*Yanxiong Li, Jiaxin Tan, Guoqing Chen, Jialong Li, Yongjie Si, Qianhua He*

School of Electronic and Information Engineering, South China University of Technology, Guangzhou, China
eeyxli@scut.edu.cn, tanjiaxin02@126.com

## Abstract

This work is an improved system that we submitted to task 1 of DCASE2023 challenge. We propose a method of low-complexity acoustic scene classification by a parallel attention-convolution network which consists of four modules, including pre-processing, fusion, global and local contextual information extraction. The proposed network is computationally efficient to capture global and local contextual information from each audio clip. In addition, we integrate other techniques into our method, such as knowledge distillation, data augmentation, and adaptive residual normalization. When evaluated on the official dataset of DCASE2023 challenge, our method obtains the highest accuracy of 56.10% with parameter number of 5.21 kilo and multiply-accumulate operations of 1.44 million. It exceeds the top two systems of DCASE2023 challenge in accuracy and complexity, and obtains state-of-the-art result. Code is at: https://github.com/Jessytan/Low-complexity-ASC.

## 1. Introduction

Acoustic scene classification (ASC) is a hot topic in the field of audio and acoustic signal processing, and many efforts are made to address the ASC problem [1]-[14]. The target of low-complexity ASC is to classify each testing clip into one of the predefined types of acoustic scenes using a lightweight model. The ASC systems are ranked by the weighted average rank of accuracy, parameter number (PN) and multiply-accumulate operations (MACs) in task 1 of DCASE2023 challenge.

To deal with the low-complexity ASC problem, almost all of existing works focused on designing a computationally-efficient network with powerful capability to distinguish different types of acoustic scenes. The methods based on convolutional neural network (CNN) and its variants are dominant solutions for the task of low-complexity ASC [12]-[21]. For example, Schmid et al designed a CNN of CP-Mobile with regularized receptive field and residual inverted bottleneck blocks [15]. Cai et al proposed a lightweight CNN of time-frequency separable convolution which was composed of a series of separable convolutional layers that exploited time and frequency domains [16]. Tan et al designed a CNN with blueprint separable convolution [17]. Madhu et al designed a residual quaternion CNN for low complexity and device-robust ASC [18]. McDonnell et al separated the ResNet into two pathways, and the separation of high and low frequencies was proved to be effective for enhancing the network's adaptability to multi-device audio clips [19].

Although the works above promoted the development of ASC, they still have shortcomings. For example, the model structures adopted in these works are almost the CNN or its variants. The model with CNN architecture is good at capturing local contextual information (LCI) from input audio clips, but it lacks the ability to effectively extract global contextual information (GCI). That is, the usage of GCI was not explicitly considered in previous works. In addition, the GCI is complementary to LCI and has been proved to be beneficial for improving the performance in other audio processing tasks. For example, a recurrent convolutional network (RCN) consisting of a bidirectional long short-term memory (BLSTM) and a convolutional block was designed to capture both GCI and LCI for obtaining better results [22]. The GCI and LCI were extracted by the BLSTM and the convolutional block, respectively. Similarly, the RCN was used for sound event detection [23] and speech enhancement [24]. The RCN in the above works still needs to be further optimized for reducing its complexity due to the following two reasons. First, the recurrent module (e.g., BLSTM) and convolutional block are connected serially and not parallelly. Second, operations in the BLSTM (recurrent module) are executed serially instead of parallelly.

In this paper, we design a parallel attention-convolution network (PACN) which is computationally efficient. In addition, we take other measures to further improve the performance of the proposed method, such as knowledge distillation (KD) [25], data augmentation (DA) [5], and adaptive residual normalization (ARN) [26]. Experimental results show that the proposed method exceeds state-of-the-art methods in accuracy and complexity. The contributions of this work are summarized as follows.

1) We design a computationally-efficient PACN to capture both GCI and LCI from input audio clips. To the best of our knowledge, the architecture of the PACN is novel and is not adopted in previous works, although each main part of the PACN consists of several typical operations.

2) We propose a low-complexity ASC method using the PACN and other techniques, including DA, KD, and ARN. We do ablation and comparison experiments on the official dataset of task 1 of DCASE2023 challenge for proving the effectiveness of our method.

## 2. Method

The framework of the proposed method is shown in Figure 1. A large-size teacher model is first trained using audio clips in the training dataset. Then, a small-size student model is generated under the guidance of the pretrained teacher model. Namely, the KD is used to train the student model which is learned from the teacher model. The DA is applied to audio clips to increase the data diversity for training both teacher and student models. In addition, the ARN is integrated into the teacher and student models for normalizing input features. The teacher and student models are designed as a PACN, while the DA, KD and ARN are similar to that used in [5], [17] and [27], respectively.

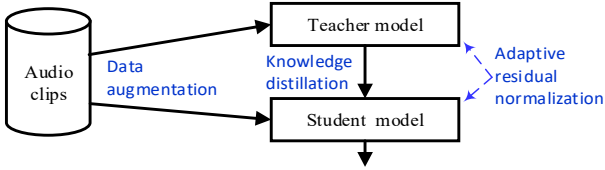

Figure 1: *The framework of the proposed method*.

### 2.1. Parallel attention-convolution network

As illustrated in Figure 2, the framework of the proposed PACN is composed of four modules, namely pre-processing, LCI extraction, GCI extraction, and fusion. The motivation for designing the proposed PACN is based on the following two considerations. First, GCI and LCI are complementary to each other and beneficial for boosting the performance of the ASC method. Hence, we design two computationally-efficient modules to explicitly capture these two kinds of information. In addition, these two modules are parallelly connected for further reducing computational complexity of the PACN. Second, a pre-processing module is designed for converting the log Mel-spectrum of each input audio clip to a transformed feature which can be effectively used to extract both GCI and LCI.

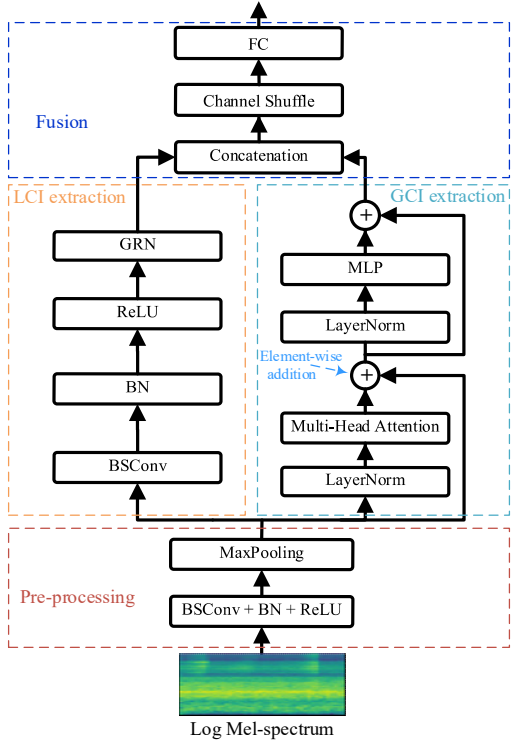

Figure 2: *The framework of the parallel attention-convolution network. BSConv: Blueprint separable convolution; ReLU: Rectified Linear Unit; BN: Batch normalization; MLP: Multi-layer perceptron; GRN: Global response normalization; FC: Fully-connected; LCI: Local contextual information; GCI: Global contextual information*.

The pre-processing module includes four classes of operations: blueprint separable convolution, rectified linear unit, batch normalization, and maximum pooling. The blueprint separable convolution is computationally efficient and consists of channel-wise convolution and point-wise convolution [28]. Compared to deep separable convolution, it focuses more on cross kernel correlation and can extract discriminative information more effectively based on the correlation within the kernel. In addition, the operation of maximum pooling is used to reduce the input feature size. Afterwards, the transformed feature is fed to two branches for extracting GCI and LCI.

The module for LCI extraction consists of four kinds of operations: blueprint separable convolution, rectified linear unit, batch normalization, and global response normalization. The first three operations are used to extract LCI, while the last operation is designed for reducing the redundancy of the transformed features. The global response normalization has three steps, including aggregation, normalization and calibration of features [29]. It normalizes the features on each channel to enhance the representational capabilities of the features.

The module for GCI extraction is composed of four kinds of operations: layer normalization, multi-head attention, element-wise addition, and multi-layer perceptron. The architecture of this module is similar to that of Transformer [30] which is good at extracting GCI. The multi-head attention is a self-attention and can capture the relationships between different audio frames. The MLP consists of two fully-connected layers and is used to further transform the feature for extracting GCI. The element-wise addition is a residual connection for element-wisely summing the output of the previous layers with the output of the current layer.

The fusion module includes three kinds of operations: concatenation, channel shuffle and fully-connected layer. This module first concatenates the GCI and LCI, and then exchanges the information of different channels by the channel shuffle and fully-connected layers.

### 2.2. Data augmentation

The diversity of training data is beneficial for improving the robustness of the proposed method. To increase the diversity of the training data, DA is applied on audio clips of training data, including mix-up [31], spectrum correction [32], pitch shift [33], and audio-mix [34].

Mix-up can make the PACN behave consistently to audio clips from various devices for improving the PACN's robustness. In each time, two batches of audio clips are loaded to memory, and the following operation is executed.

$$X' = \eta X_i + (1-\eta) X_j, \quad (1)$$

where $X_i$ and $X_j$ denote the two batches of audio clips and $\eta$ follows the Beta distribution. By loading two batches of audio clips, more information can be utilized during mixing process. Hence the PACN's performance is expected to be improved.

Spectrum correction aims to convert the input spectrum to a corrected spectrum by a reference device. It is executed by two steps. In the first step, a correction coefficient is generated by calculating the mean of several pairs of aligned spectra. The correction coefficient of device $A$ to the reference device is set to the ratio of the frequency response of the reference device to the counterpart of device $A$. The reference device's response is defined as the mean of the counterparts of multiple devices. In the second step, the corrected spectrum of device $A$ is obtained by multiplying the correction coefficient with the original spectrum of the audio clips acquired by device $A$.

Pitch shift is to resample the original audio clips at different sampling frequencies with a specific step size. The pitches of audio clips obtained by different devices are generally different to each other. Pitch shifting is executed before the extraction of log Mel-spectrogram.

Audio-mix is to mix two audio clips. The audio clips are stochastically chosen from the same class of acoustic scenes for simulating more recording devices, smoothing transition and reducing differences among various recording devices.

### 2.3. Knowledge distillation

In DCASE2023 challenge, the complexity of teacher model is not used for performance evaluation, but the complexity of student model is adopted to rank the complexities of different methods. In this paper, we use the KD technique to transfer knowledge of the pre-trained large-scale teacher model to the small-scale student model. As a result, the performance of the proposed method is expected to be improved. The loss function for KD is defined by

$$\ell = \lambda \ell_h \big( f(z_s), y \big) + (1-\lambda) \ell_{kd} \left( f\left(\frac{z_s}{T}\right), f\left(\frac{z_t}{T}\right) \right), \quad (2)$$

where $\lambda$ and $T$ denote a weight coefficient and temperature coefficient, respectively; $\ell_h$ and $\ell_{kd}$ represent hard-label loss and distillation loss, respectively; $z_s$ and $z_t$ stand for the logits of student and teacher models, respectively.

### 2.4. Adaptive residual normalization

Frequency instance normalization (FIN) [26] generalizes the features in the device domain by applying instance normalization along the frequency dimension, and is defined by

$$FIN(x) = \frac{x - \mu_{nf}}{\sqrt{\sigma_{nf}^2 + \epsilon}}, \quad (3)$$

where $x \in \mathbb{R}^{n \times c \times f \times t}$ represents a 4-dimensional input feature vector with batch size, channels, frequency bins and temporal clips; $\mu_{nf}$ and $\sigma_{nf}$ represent the mean and standard deviation of input feature vector along the $n$ and $f$ dimensions, respectively; $\epsilon$ stands for a minimum constant coefficient for numerical stability.

Inspired by the success in [27], we introduce ARN by adding trainable parameters for controlling the trade-off between device identity and FIN. ARN is defined by

$$ARN(x) = \big(\rho \times x + (1-\rho) \times FIN(x)\big) \times \gamma + \beta, \quad (4)$$

where $\rho$, $\gamma$ and $\beta$ represent trainable parameters for balancing, scaling and shifting. ARN is inserted after the first convolution layer and each stage of BSConv blocks. In summary, ARN is a normalization module with adaptive parameters, which enables the network to normalize the input data based on the specific task and the distribution of input data. As a result, the performance of the network can be improved by providing the network with more flexibility in handling different types of input data.

## 3. Experiments

In this section, we will describe experimental data, setup, results, and corresponding discussions in detail.

### 3.1. Experimental data

We conduct our experiments on the TAU Urban Acoustic Scene 2022 Mobile development dataset (TAU22) which consists of audio clips acquired by mobile devices in urban environments. The dataset includes 230,350 audio clips. Each clip is with a duration of 1 second and a hard label of an acoustic scene. The dataset contains audio clips from 10 cities and 9 devices: 3 real devices (A, B, C) and 6 simulated devices (S1-S6). Audio clips recorded by devices B, C, and S1-S6 are composed of audio segments that are randomly selected from simulated recordings. Hence, all of these audio clips overlap with the audio clips from device A, but not necessarily with each other. The total amount of audio clips in the development dataset is 64 hours. There are 10 classes of acoustic scenes, including Airport, Metro station, Indoor shopping mall, Pedestrian street, Public square, Street with medium level of traffic, Travelling by a tram, Travelling by a bus, Travelling by an underground metro, and Urban park. The data usage in our experiments is the same to that in task 1 of DCASE2023 challenge. The detailed information of the data partitions is described in the official website of DCASE2023 challenge [35].

### 3.2. Experimental setup

Audio clips are split into frames via a Hamming window whose length is 4096 with 1/6 overlapping. Short-time Fourier transform is then performed on each frame for obtaining linear power spectrum which is smoothed with a bank of triangular filters for extracting log Mel-spectrum. In addition, the delta coefficients of log Mel-spectrum are calculated and then concatenated with the log Mel-spectrum to form the input audio feature. The final size of input audio feature is: 256×65×2, where 256, 65 and 2 denote numbers of frequency-band, frame and channel, respectively.

We train the models for 100 epochs using the Adam optimizer [36] with batch size of 16. The learning rate is scheduled to linearly increase from 0 to 0.002 in ten epochs as a warmup, then decay to 0 with cosine annealing for the rest of epochs. The coefficients for DA and KD are set as follows: α=0.4, T=2, λ=0.226. The initial values of $\rho$, $\gamma$ and $\beta$ are set to 0.5, 1.0 and 0.0, respectively; and their values are adaptively adjusted during the training process. Accuracy, PN and MACs are used as the performance metrics.

### 3.3. Experimental results

The effectiveness of DA, KD and ARN for improving performance of ASC methods are proved in previous works of [5], [17] and [27], respectively. Hence, their impacts on the proposed method are not discussed in this paper. Instead, we present the impacts of modules for extracting GCI and LCI on the performance of our method, as given in Table 1.

Table 1: *Impacts of GCI and LCI on our method's performance.*

| Methods | Acc (%) | PN (kilo) | MACs (million) |
| --- | --- | --- | --- |
| with (GCI//LCI) | 56.10 | 5.21 | 1.44 |
| with (GCI+LCI) | 54.73 | 5.37 | 1.46 |
| without fusion | 54.62 | 5.38 | 1.46 |

(GCI//LCI) and (GCI+LCI) denote parallel and series connections of GCI and LCI extraction modules, respectively. Acc: Accuracy.

When the extraction modules of LCI and GCI are connected parallelly, the proposed method obtains the accuracy of 56.10%, PN of 5.21 kilo and MACs of 1.44 million. The proposed method obtains the lowest accuracy of 54.62% when the LCI and GCI are not fused. In addition, compared to the case of (GCI+LCI), our method achieves higher accuracy and lower PN and MACs in the case of (GCI//LCI).

In the comparison experiment, we compare the proposed method with the top two systems and baseline system submitted to task 1 of DCASE2023 challenge. Table 2 shows the results obtained by different methods when they are evaluated on the development dataset of task 1 of DCASE2023 challenge. The proposed method achieves accuracy of 56.10%, PN of 5.21 kilo, and MACs of 1.44 million. The accuracy obtained by our method is higher than that of three comparative methods. Both PN and MACs of the proposed method are lower than the counterparts of three comparative methods. That is, the proposed method outperforms the top two systems and baseline system of DCASE2023 challenge in all performance metrics (accuracy and complexity).

Table 2: *Results obtained by different methods*.

| Methods | Acc (%) | PN (kilo) | MACs (million) |
|---|---|---|---|
| Ours | 56.10 | 5.21 | 1.44 |
| CP-Mobile [15] | 54.66 | 5.77 | 1.58 |
| TF-SpeNet [16] | 53.90 | 6.83 | 1.65 |
| DCASE baseline | 42.90 | 46.51 | 29.23 |

It should be noted that the most important goal of this work is to obtain an ASC system with higher accuracy and minimum complexity. Therefore, we compare the two top systems with lowest complexity submitted to the DCASE2023 challenge with the aim of obtaining an ASC system with low-complexity and high-accuracy. The ultimate goal of this work is to deploy a high-accuracy ASC system on intelligent audio terminals.

In the aforementioned experiments, the recording devices of testing clips are the same as that of training clips. Table 3 presents the accuracies obtained by different methods when the recording devices of testing clips are unseen in training clips. It can be seen from Table 3 that the proposed method achieves the highest accuracy score of 52.74%. This result indicates that our method has stronger robustness for different recording devices compared to other two methods.

Table 3: *Accuracies obtained by different methods on testing audio clips acquired by unseen devices.*

| Methods | Accuracy (%) |
|---|---|
| Ours | 52.74 |
| CP-Mobile [15] | 51.83 |
| TF-SpeNet [16] | 50.90 |

In addition, we conduct significance test to verify whether accuracy differences between different methods are statistically significant. We do not know the distribution of samples, nor do we make any assumptions about the distribution of samples. Hence, non-parametric instead of parametric significance test is used. Specifically, we adopt the Friedman test (null-hypothesis test) with the Nemenyi test (post-hoc test) [37], which has been widely applied for significance test. We evaluate four methods separately on each of twenty data subsets which are randomly extracted from experimental dataset. Four methods are ranked on each data subset based on their accuracy scores. This procedure yields a total of twenty rankings for four methods. Figure 3 shows results of significance test for various methods. We plot the number of times each method obtains each rank in Figure 3 (a). The methods are ordered by their average ranks (decimals in square brackets). Our method (Ours) is the best method, and ranks first on twenty of the twenty data subsets with an average rank of 1.0. CP-Mobile, TF-SpeNet and DCASE baseline rank second, third and fourth with an average rank of 2.1, 2.9 and 4.0, respectively. Figure 3 (b) gives results of significance test with a confidence level of $\alpha=0.05$. The horizontal axis gives average rank of each method, black-thick crossbars link methods for which there is insufficient evidence to declare them statistically significantly different. Our method significantly exceeds two methods (TF-SpeNet and DCASE baseline), since our method is not linked to them. However, the difference between our method and the CP-Mobile is not significant, since they are linked by one black-thick crossbar.

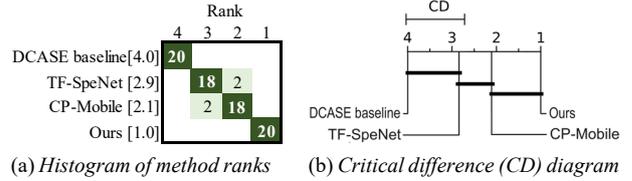

(a) *Histogram of method ranks*    (b) *Critical difference (CD) diagram*

Figure 3: *Results of significance test*. (a) *Visualization of the number of times each method achieves each rank*. (b) *Visualization of significance differences between different methods*.

To know the detailed confusions among different classes of acoustic scenes, Figure 4 illustrates the confusion matrix produced by the proposed method. The accuracy scores on the main diagonal of the confusion matrix is significantly higher than that at other positions. This result indicates that the accuracy socre of correct classification for each category of acoustic scenes is higher than the accuracy score of confusion between categories. The accuracies for Public square, Metro station, and Street pedestrian are lower than that for other acoustic scenes. The reason may be that these three acoustic scenes have larger intra-class differences of time-frequency properties. In addition, the confusions between Metro and Tram are larger than that between other pairs of acoustic scenes. The possible reason is that the time-frequency properties of these two classes of acoustic scenes are more similar than others. Therefore, they are more prone to mutual confusion.

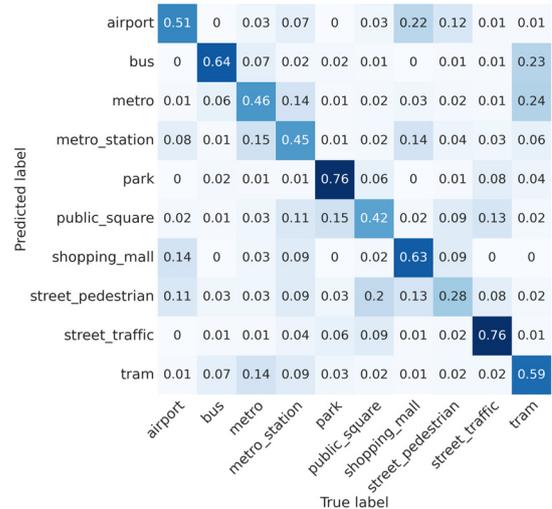

Figure 4: *Confusion matrix of the proposed method.*

## 4. Conclusions

In this paper, we designed a model of PACN which is novel in architecture. In addition, we proposed a low-complexity ASC method using the PACN. When evaluated on the official dataset of task 1 of DCASE2023 challenge, our method exceeded the top two systems and baseline systme submitted to task 1 of DCASE2023 challenge in all performance metrics.

Though our method achieves state-of-the-art performance, it still needs to be improved. For example, only self-attention is used in the PACN without considering mutual-attention. The latter should be useful for performance improvement. In addition, we verified the effectiveness of the proposed method by preliminary experiments due to the limitation of the paper length. Next, we will perform comprehensive experiments to discuss the performance of our method. In addition, We will consider making our method more lightweight and deploying it on intelligent audio terminals.

## 5. Acknowledgements

This work was partly supported by the national natural science foundation of China (62371195, 62111530145, 61771200), international scientific research collaboration project of Guangdong (2023A0505050116), Guangdong basic and applied basic research foundation (2022A1515011687), and Guangdong provincial key laboratory of human digital twin (2022B1212010004).